# Lattice reduction by cubification

Authors

**Cyril Cayron[a]***

[a]Laboratory of Thermo Mechanical Metallurgy (LMTM), PX Group Chair, EPFL, Rue de la Maladière 71b, Neuchâtel, 2000, Switzerland

Correspondence email: cyril.cayron@epfl.ch

**Synopsis** An algorithm of lattice reduction based directional and hyperplanar projections and driven by the reduction of the lattice rhombicity is proposed. It may have an impact beyond crystallography.

**Abstract** Lattice reduction is a NP-hard problem well known in computer science and cryptography. The Lenstra-Lenstra-Lovász (LLL) algorithm based on the calculation of orthogonal Gram-Schmidt (GS) bases is efficient and gives a good solution in polynomial time. Here, we present a new approach called cubification that does not require the calculation of the GS bases. It relies on complementary directional and hyperplanar reductions. The deviation from cubicity at each step of the reduction process is evaluated by a parameter called lattice rhombicity, which is simply the sum of the absolute values of the metric tensor. Cubification seems to equal LLL; it even outperforms it in the reduction of columnar matrices. We wrote a Python program that is ten times faster than a reference Python LLL code. This work may open new perspectives for lattice reduction and may have implications and applications beyond crystallography.

## 1. Introduction

In computer science, digital communication and cryptography, lattice problems form a class of intricate NP-hard (non-deterministic polynomial-time) optimization problems, such as the shortest vector problem, the closest vector problem, and the shortest basis problem. Solving one of them would help to solve the others. Lattice reduction algorithms aim at solving the last problem. Given a lattice $\mathcal{L}$ made of $N$ free vectors $\mathbf{b}_i$, the algorithm should give new relatively short, nearly orthogonal vectors $\mathbf{b}'_i$ of the same lattice $\mathcal{L}$. It consists in finding integers $z_{ij}$ such that $\mathbf{b}_i = \sum_{j=1}^{N} z_{ij}\,\mathbf{b}'_j$ and $\mathcal{L} = \{\mathbb{Z}\,\mathbf{b}_i\} = \{\mathbb{Z}\,\mathbf{b}'_j\}$, where the $\{\mathbb{Z}\ .\}$ means all the linear combinations with integer coefficients. The number of vectors cannot be larger the space dimension; often, both are the same. The set of vectors are linked by an integer matrix $\mathbf{Z}$ of determinant ±1. This relation can be written $\begin{bmatrix} \mathbf{b}_1 \\ \vdots \\ \mathbf{b}_i \\ \vdots \\ \mathbf{b}_N \end{bmatrix} = \mathbf{Z} \begin{bmatrix} \mathbf{b}'_1 \\ \vdots \\ \mathbf{b}'_j \\ \vdots \\ \mathbf{b}'_N \end{bmatrix}$ with $\mathbf{Z} \in \mathbb{Z}^{N.N}$ and $Det(\mathbf{Z}) = \pm 1$, where $\mathbf{b}_i$ and $\mathbf{b}'_j$ refer to the vectors themselves, not to their coordinates.

The most popular algorithm to tackle the lattice reduction problem was proposed nearly forty years ago by Lenstra-Lenstra-Lovász (1982) and is still considered as the main reference in the domain. It is so important that a complete book was devoted to it (Nguyen & Vallée, 2010). The reader can also consult Wikipedia (2020). We just give here some key points. One is the Gram-Schmidt orthogonalization routine in which one associates to any basis $\{\mathbf{b}_1, \dots, \mathbf{b}_k, \dots, \mathbf{b}_m\}$ , $m \leq N$, an orthogonal basis $\{\mathbf{b}^*_1, \dots, \mathbf{b}^*_k, \dots, \mathbf{b}^*_m\}$ by a series of projections $\mathbf{b}^*_k = \mathbf{b}_k - \sum_{i<k} u_{i,k} \mathbf{b}_i$ with $u_{i,k} = \frac{\mathbf{b}_k \cdot \mathbf{b}^*_i}{\mathbf{b}^*_i \cdot \mathbf{b}^*_i}$. The vectors $\mathbf{b}^*_k$ are not anymore integer ("reticular" in crystallographic language); they remain however rational. The LLL algorithm works in two steps repeated iteratively. The first step is the quasi orthogonalization using values of the coefficients $u_{i,k}$. The vectors $\mathbf{b}_i$ are replaced by $\mathbf{b}_i - \lfloor u_{i,k} \rceil \, \mathbf{b}_k$ , for $k$ between 1 and $i - 1$, where $\lfloor u_{i,k} \rceil$ means the nearest integer of $u_{i,k}$. The Gram-Schmidt basis should be actualized during the process. The second step relies on a criterion to determine whether or not the vectors $\mathbf{b}_i$ and $\mathbf{b}_i$ should be swapped: the swap is made when $\left\|\mathbf{b}^*_{i+1} + u_{i,i+1}\mathbf{b}^*_i\right\|^2 < \alpha \|\mathbf{b}^*_i\|^2$, where $\alpha$ is a constant arbitrarily chosen between ¼ and 1 and fixed once for all. Often, the value $\alpha = \frac{3}{4}$ is chosen. It is usual in lattice reduction problem to present the vectors as rows to form a matrix. In crystallography, we generally write the coordinates in columns and keep the row notation for planes, i.e. for vectors of the reciprocal space. In order to avoid any confusion, we will write $\mathbf{b}^{\mathrm{t}}$ a vector $\mathbf{b}$ written in row, where "t" means "transpose". In a space of dimension *N*, a vector $\mathbf{b}$ is a $N \times 1$ matrix, and $\mathbf{b}^{\mathrm{t}}$ a $1 \times N$ matrix. In this paper, the scalar product of two vectors $\mathbf{p}$ and $\mathbf{u}$ is noted equivalently $\mathbf{p} \cdot \mathbf{u} = \mathbf{p}^{\mathrm{t}} \, \mathbf{u}$, but the last notation will be preferred in order to keep the distinction between row and column vectors, between reciprocal and direct space vectors.

A typical low dimensional example of lattice reduction is the set of three vectors in 3 dimensions, $\mathbf{b}^{\mathrm{t}}_1 = [1,1,1]$, $\mathbf{b}^{\mathrm{t}}_2 = [-1,0,2]$, and $\mathbf{b}^{\mathrm{t}}_3 = [3,5,6]$. They form the matrix $\mathbf{B} = \begin{bmatrix} 1 & 1 & 1 \\ -1 & 0 & 2 \\ 3 & 5 & 6 \end{bmatrix}$. The reduced lattice is the matrix $\mathbf{B}' = \begin{bmatrix} 0 & 1 & 0 \\ 1 & 0 & 1 \\ -1 & 0 & 2 \end{bmatrix}$. One can check that $\mathbf{b}_1 = \mathbf{b}'_1 + \mathbf{b}'_2$, $\mathbf{b}_2 = \mathbf{b}'_3$, and $\mathbf{b}_3 = 5\,\mathbf{b}'_1 + 4\,\mathbf{b}'_2 + \mathbf{b}'_3$ . The "t" is not indicated because these linear equations can written equivalently in rows or columns. The integer coefficients of linearity could be found by calculating $\mathbf{Z} = \mathbf{B}\,(\mathbf{B}')^{-1}$.

I came to be interested in lattice reduction by a crystallographic study (Cayron, 2021b) in which a method is proposed to determine a unit cell attached to a hyperplane of dimension $N-1$ given by a vector of the reciprocal lattice $\mathbf{p}^{\mathrm{t}}$. The unit cell is made of $N-1$ vectors $\{\mathbf{b}_2, \dots, \mathbf{b}_i, \dots, \mathbf{b}_N\}$ inside the plane, i.e. such that $\mathbf{p}^{\mathrm{t}}\,\mathbf{b}_i = 0$, and one vector $\mathbf{b}_1$ in the first layer, i.e. such that $\mathbf{p}^{\mathrm{t}}\,\mathbf{b}_1 = 1$. The initial result obtained in (Cayron, 2021) was not yet fully satisfying because the vectors $\{\mathbf{b}_1, \dots, \mathbf{b}_i, \dots, \mathbf{b}_N\}$ forming the unit cell were not yet reduced, i.e. other unit cells of same volume based on smaller vectors $\{\mathbf{b}'_1, \dots, \mathbf{b}'_i, \dots, \mathbf{b}'_N\}$ were clearly possible for the same hyperplane. How to determine them? This question made us realize that lattice reduction is not as simple as one could think and that LLL could be a good option. However, despite its highly recognized and well-establish status in computer science, some points of the LLL algorithm may seem strange for a crystallographer. First, the calculations in LLL use GS orthogonalized bases, called in crystallography "structure tensors", but we know that calculations are more effectively performed with the metric tensors

$$\boldsymbol{\mathcal{M}} = \begin{bmatrix} \mathbf{b}_1^{\mathrm{t}} \\ \vdots \\ \mathbf{b}_i^{\mathrm{t}} \\ \vdots \\ \mathbf{b}_N^{\mathrm{t}} \end{bmatrix} (\mathbf{b}_1, \dots, \mathbf{b}_j, \dots, \mathbf{b}_N) = \begin{bmatrix} \mathbf{b}_1^2 & \dots & \mathbf{b}_1^{\mathrm{t}}\,\mathbf{b}_j & \dots & \mathbf{b}_1^{\mathrm{t}}\,\mathbf{b}_N \\ \vdots & \dots & \vdots & \dots & \vdots \\ \mathbf{b}_i^{\mathrm{t}}\,\mathbf{b}_1 & \dots & \mathbf{b}_i^{\mathrm{t}}\,\mathbf{b}_j & \dots & \mathbf{b}_i^{\mathrm{t}}\,\mathbf{b}_N \\ \vdots & \dots & \vdots & \dots & \vdots \\ \mathbf{b}_N^{\mathrm{t}}\,\mathbf{b}_1 & \dots & \mathbf{b}_N^{\mathrm{t}}\,\mathbf{b}_j & \dots & \mathbf{b}_N^2 \end{bmatrix}$$

. Second, the constant α and the way it should be "arbitrarily" chosen is not explicit. It is like a strength of convergence in the algorithm; greater values lead to stronger reductions; it has an effect on the final norms of the reduced vectors, and more precisely it permits to bound the product of the squared norms $\prod_{i=1}^{N} \|\mathbf{b}_i\|^2$. This means that even if LLL uses projected vectors and orthogonalizations, the convergence criterion is mainly established on the norms, as if minimizing the norms were equivalent to orthogonalizing the basis. This assumption seems "reasonable" but has never been demonstrated, and it may be actually false. For example, the reader can look at the $20 \times 20$ matrix in Figure 1a representing the list of 20 vectors written in rows. After LLL reduction (Figure 1b) we can notice that the first coordinates (at the left) are highly occupied with ±1 whereas the last coordinates (at the right) are nearly all null, showing an unbalanced distribution of the coordinates, and thus an excess of "rhombicity". Consequently, a more general criterion to estimate how much "small and cubic" is a basis should be introduced. There is probably not only a unique way to define it. In the present work, we propose for a basis $\{\mathbf{b}_1, \dots, \mathbf{b}_i, \dots, \mathbf{b}_N\}$ the simple "crystallographic" parameter based on the metric tensor

$$R = \sum_{i,j} \left|\boldsymbol{\mathcal{M}}_{ij}\right| = \sum_{i \le N} \|\mathbf{b}_i\|^2 + 2 \sum_{i<j\le N} \mathbf{b}_i^{\mathrm{t}}\,\mathbf{b}_j \qquad (1)$$

We will call it "lattice rhombicity". This parameter contains the information on both the norms and the orthogonalities between the vectors. A lowest "lattice rhombicity" indicates a smaller and more cubic basis. Please note that the term "rhombicity" as a specific meaning in a branch of mathematics for symmetric second-rank tensors in three-dimensional Euclidean space, but that given in the present paper; that is why we specify "lattice rhombicity". If one wants to evaluate only the norms, we propose to use the parameter $S = \sum_{i \leq N} ||\mathbf{b}_i||^2$.

(a) Input (the coordinates of the vectors are written in rows):

$$
\begin{bmatrix}
1 & 0 & 0 & 0 & 0 & 0 & 0 & 0 & 0 & 0 & 0 & 0 & 0 & 0 & 0 & 0 & 0 & 0 & 0 & 75 \\
0 & 1 & 0 & 0 & 0 & 0 & 0 & 0 & 0 & 0 & 0 & 0 & 0 & 0 & 0 & 0 & 0 & 0 & 0 & 436 \\
0 & 0 & 1 & 0 & 0 & 0 & 0 & 0 & 0 & 0 & 0 & 0 & 0 & 0 & 0 & 0 & 0 & 0 & 0 & 1586 \\
0 & 0 & 0 & 1 & 0 & 0 & 0 & 0 & 0 & 0 & 0 & 0 & 0 & 0 & 0 & 0 & 0 & 0 & 0 & 1030 \\
0 & 0 & 0 & 0 & 1 & 0 & 0 & 0 & 0 & 0 & 0 & 0 & 0 & 0 & 0 & 0 & 0 & 0 & 0 & 1921 \\
0 & 0 & 0 & 0 & 0 & 1 & 0 & 0 & 0 & 0 & 0 & 0 & 0 & 0 & 0 & 0 & 0 & 0 & 0 & 569 \\
0 & 0 & 0 & 0 & 0 & 0 & 1 & 0 & 0 & 0 & 0 & 0 & 0 & 0 & 0 & 0 & 0 & 0 & 0 & -721 \\
0 & 0 & 0 & 0 & 0 & 0 & 0 & 1 & 0 & 0 & 0 & 0 & 0 & 0 & 0 & 0 & 0 & 0 & 0 & 1183 \\
0 & 0 & 0 & 0 & 0 & 0 & 0 & 0 & 1 & 0 & 0 & 0 & 0 & 0 & 0 & 0 & 0 & 0 & 0 & 1570 \\
0 & 0 & 0 & 0 & 0 & 0 & 0 & 0 & 0 & 1 & 0 & 0 & 0 & 0 & 0 & 0 & 0 & 0 & 0 & -6665 \\
0 & 0 & 0 & 0 & 0 & 0 & 0 & 0 & 0 & 0 & 1 & 0 & 0 & 0 & 0 & 0 & 0 & 0 & 0 & 123 \\
0 & 0 & 0 & 0 & 0 & 0 & 0 & 0 & 0 & 0 & 0 & 1 & 0 & 0 & 0 & 0 & 0 & 0 & 0 & 890 \\
0 & 0 & 0 & 0 & 0 & 0 & 0 & 0 & 0 & 0 & 0 & 0 & 1 & 0 & 0 & 0 & 0 & 0 & 0 & 6 \\
0 & 0 & 0 & 0 & 0 & 0 & 0 & 0 & 0 & 0 & 0 & 0 & 0 & 1 & 0 & 0 & 0 & 0 & 0 & 742 \\
0 & 0 & 0 & 0 & 0 & 0 & 0 & 0 & 0 & 0 & 0 & 0 & 0 & 0 & 1 & 0 & 0 & 0 & 0 & 33 \\
0 & 0 & 0 & 0 & 0 & 0 & 0 & 0 & 0 & 0 & 0 & 0 & 0 & 0 & 0 & 1 & 0 & 0 & 0 & 888 \\
0 & 0 & 0 & 0 & 0 & 0 & 0 & 0 & 0 & 0 & 0 & 0 & 0 & 0 & 0 & 0 & 1 & 0 & 0 & 14 \\
0 & 0 & 0 & 0 & 0 & 0 & 0 & 0 & 0 & 0 & 0 & 0 & 0 & 0 & 0 & 0 & 0 & 1 & 0 & 769 \\
0 & 0 & 0 & 0 & 0 & 0 & 0 & 0 & 0 & 0 & 0 & 0 & 0 & 0 & 0 & 0 & 0 & 0 & 1 & 1234 \\
0 & 0 & 0 & 0 & 0 & 0 & 0 & 0 & 0 & 0 & 0 & 0 & 0 & 0 & 0 & 0 & 0 & 0 & 0 & -852
\end{bmatrix}
$$

(b) LLL output:

$$
\begin{bmatrix}
-1 & 0 & -1 & 1 & 1 & -1 & 1 & 0 & 0 & 0 & 0 & 0 & 0 & 0 & 0 & 0 & 0 & 0 & 0 & 0 \\
0 & 0 & 0 & -1 & 0 & 1 & 1 & 1 & 0 & 0 & 0 & 0 & 0 & 0 & 0 & 0 & 0 & 0 & 0 & 1 \\
1 & 0 & 0 & -1 & 0 & 1 & 0 & -1 & 1 & 0 & 0 & 0 & 0 & 0 & 0 & 0 & 0 & 0 & 0 & 1 \\
0 & 0 & 0 & 1 & -1 & 0 & 0 & 0 & 0 & 0 & 0 & 1 & 0 & 0 & 0 & 0 & 0 & 0 & 0 & -1 \\
0 & 0 & -1 & 0 & 0 & -1 & 1 & 1 & 1 & 0 & 1 & 0 & 0 & 0 & 0 & 0 & 0 & 0 & 0 & 0 \\
0 & 0 & 0 & 1 & -1 & -1 & 0 & 0 & 1 & 0 & 0 & 0 & 0 & 1 & 0 & 0 & 0 & 0 & 0 & 0 \\
0 & 0 & 0 & 0 & 1 & -1 & 1 & 1 & 0 & 0 & 0 & 0 & 0 & 1 & 0 & 0 & 0 & 0 & 0 & 0 \\
-1 & 0 & 0 & 0 & 0 & -1 & 0 & -1 & 0 & 0 & 1 & 0 & 0 & 0 & 0 & 0 & 0 & 0 & 0 & 0 \\
0 & 0 & 0 & -1 & 0 & 0 & 0 & 0 & -1 & 0 & 0 & 1 & 1 & 0 & 0 & 0 & 0 & 0 & 0 & 0 \\
0 & -1 & 0 & 0 & 0 & 1 & 0 & 0 & 1 & 0 & 0 & 0 & 0 & 0 & 0 & 0 & 0 & 0 & 0 & -1 \\
0 & -1 & 1 & 0 & 0 & 0 & 0 & -1 & 0 & 0 & 0 & 0 & 0 & 0 & 1 & 0 & 0 & 0 & 0 & 0 \\
0 & -1 & 0 & 1 & 0 & 0 & 0 & 0 & -1 & 0 & 1 & 0 & 0 & 0 & 0 & 0 & 0 & 0 & 0 & -1 \\
-1 & 0 & 0 & 0 & 0 & 0 & 0 & 0 & 0 & 0 & 0 & 1 & 0 & 0 & 0 & 1 & 0 & 0 & 0 & -1 \\
0 & 1 & 0 & 0 & 0 & 0 & 0 & 1 & 0 & 0 & 1 & -1 & 0 & 0 & 0 & 0 & 0 & 0 & 0 & 0 \\
0 & 0 & 0 & -1 & 0 & 1 & -1 & -1 & 0 & 0 & 0 & 1 & 0 & 0 & 1 & 0 & 0 & 0 & 0 & 0 \\
-1 & 0 & 0 & 1 & 0 & 0 & 0 & 0 & 0 & 0 & -1 & 0 & 1 & 0 & 0 & 0 & 1 & 0 & 0 & 0 \\
0 & 0 & 0 & 0 & -1 & 0 & -1 & -1 & 0 & 0 & -1 & 0 & 0 & 0 & 1 & 0 & 0 & 1 & 0 & 0 \\
-1 & 0 & -1 & 0 & -1 & -1 & 0 & 0 & 0 & 0 & 0 & 0 & 0 & 1 & 0 & 0 & 0 & 0 & 0 & -1 \\
0 & 0 & 0 & 0 & 1 & 0 & 1 & 0 & 0 & 0 & 1 & 0 & 0 & 0 & 0 & 0 & 0 & 0 & 1 & 1 \\
-1 & 0 & 0 & 1 & 0 & 0 & -1 & 0 & 0 & 1 & -1 & 0 & 0 & 0 & 0 & 0 & 0 & 0 & 0 & 0
\end{bmatrix}
$$

**Figure 1** Example of the LLL algorithm with $20 \times 20$ matrix representing a list of 20 vectors whose coordinates are written in rows. (a) Input list. (b) Output list determined with the function *LatticeReduce* of *Mathematica*. Note that the coordinates in output matrix are still unbalanced. The coordinates in columns13 to 19 are nearly "empty", they contain only one or two ±1, whereas the columns 1 to 12 and 20 contain more than four ±1. Before reduction, in (a), the values of the rhombicity $R = \sum_N ||\mathbf{b}_i||^2 + 2\sum_{i<j\leq N} \mathbf{b}_i^{\mathrm{t}} \mathbf{b}_j$ and of the sum of the squares of the norms $S = \sum_N ||\mathbf{b}_i||^2$are $R = 453988268$, $S = 61580172$. After reduction, in (b), they become $R = 531$, $S = 99$.

Beside our work on the unit cells linked to hyperplanes (Cayron, 2021b), we thought that some of the mathematical tools used in our research on twins (Cayron, 2020, 2021a), such as the dimension reduction by the use of left inverses could also be applied to the lattice reduction problem. This lead us to attempt a first method called "hyperplanar reduction". The results were good but not as good as with LLL. It quickly appeared that the process would work better if the basis $\{\mathbf{b}_2, \ldots, \mathbf{b}_i, \ldots, \mathbf{b}_N\}$ associated with the hyperplane could be already reduced, even if only partially. Thus, an additional process called "directional reduction" was introduced. Neither directional reduction nor hyperplanar reduction can reach the performance of the LLL algorithm, but working together, they equal and sometimes outperform it.

The principle of directional reduction will be presented in §2. It helps to obtain a reduced lattice with significantly lower $R$ and $S$ values. They remain however higher than with LLL. The hyperplanar reduction will be explained in §3; it continues to decrease $R$ and $S$, and improves the cubicity. In §4, it will be shown how cycling directional and hyperplanar reductions permits to obtain $R$ and $S$ comparable, and sometimes better than with LLL.

## 2. Directional reduction

### 2.1. Lagrange's division

We consider a basis in *N* dimensions made of *N* integer vectors $\{\mathbf{b}_1, \ldots, \mathbf{b}_i, \ldots, \mathbf{b}_N\}$ initially sorted by norms, from the lowest to the highest norms, i.e. such that $\|\mathbf{b}_1\| \leq \cdots \leq \|\mathbf{b}_i\| \leq \|\mathbf{b}_{i+1}\| \ldots \leq \|\mathbf{b}_N\|$.

We consider two vectors $\mathbf{b}_i$ and $\mathbf{b}_j$ in the list such that $\|\mathbf{b}_i\| \leq \|\mathbf{b}_j\|$. The vector $q\mathbf{b}_i$ with $q = \frac{\mathbf{b}_i^t \mathbf{b}_j}{\mathbf{b}_i^t \mathbf{b}_i}$ is the orthogonal projection of $\mathbf{b}_j$ on $\mathbf{b}_i$. It not an integer vector but remains however a rational. It can be approximated by the integer vector $\lfloor q \rceil \mathbf{b}_i$ , where $\lfloor q \rceil$ is the integer closest to $q$ and computed by $\lfloor q \rceil = int(round(q))$. If the coordinates of the vectors are such that $|q| \geq \frac{1}{2}$, i.e. $\lfloor q \rceil \neq 0$, the reduced vector $\mathbf{r} = \mathbf{b}_j - \lfloor q \rceil \mathbf{b}_i$ belongs to the lattice generated by $\mathbf{b}_i$ and $\mathbf{b}_j$ and its norm is such that $\|\mathbf{r}\| \leq \|\mathbf{b}_j\|$. In the limit case $|q| = \frac{1}{2}$, the triangle made by $(\mathbf{b}_i, \mathbf{b}_j, \mathbf{r})$ is isosceles, i.e. $\|\mathbf{r}\| = \|\mathbf{b}_j\|$. Note that in some cases, the norm of $\mathbf{r}$ may even be lower than that of $\mathbf{b}_i$. Two cases should be distinguished in the algorithm: if $\lfloor q \rceil = 0$, nothing changes and a next pair of vectors should be considered; if $\lfloor q \rceil \neq 0$, a change should be done, and two algorithm variants are proposed:

> <u>Variant *Append*</u>: the vectors $\mathbf{b}_i$ and $\mathbf{b}_j$ are deleted from the list, and the vectors $\mathbf{r}$ and $\mathbf{b}_i$ are appended at the end of the list.
>
> <u>Variant *Insert*</u>: if $\|\mathbf{r}\| \leq \|\mathbf{b}_i\|$, $\mathbf{r}$ replaces $\mathbf{b}_i$ and $\mathbf{b}_i$ replaces $\mathbf{b}_j$ in the list; else, $\mathbf{r}$ replaces $\mathbf{b}_j$ in the list.

The process is recursively repeated without sorting the list of vectors until all the values $\lfloor q \rceil$ become null for all the pairs of vectors in the basis. The method is quite similar to Lagrange's division described by Nguyen & Vallée (2010), except that for a pair of vectors $\mathbf{b}_i$ and $\mathbf{b}_j$ the division is made only once and not repeated recursively.

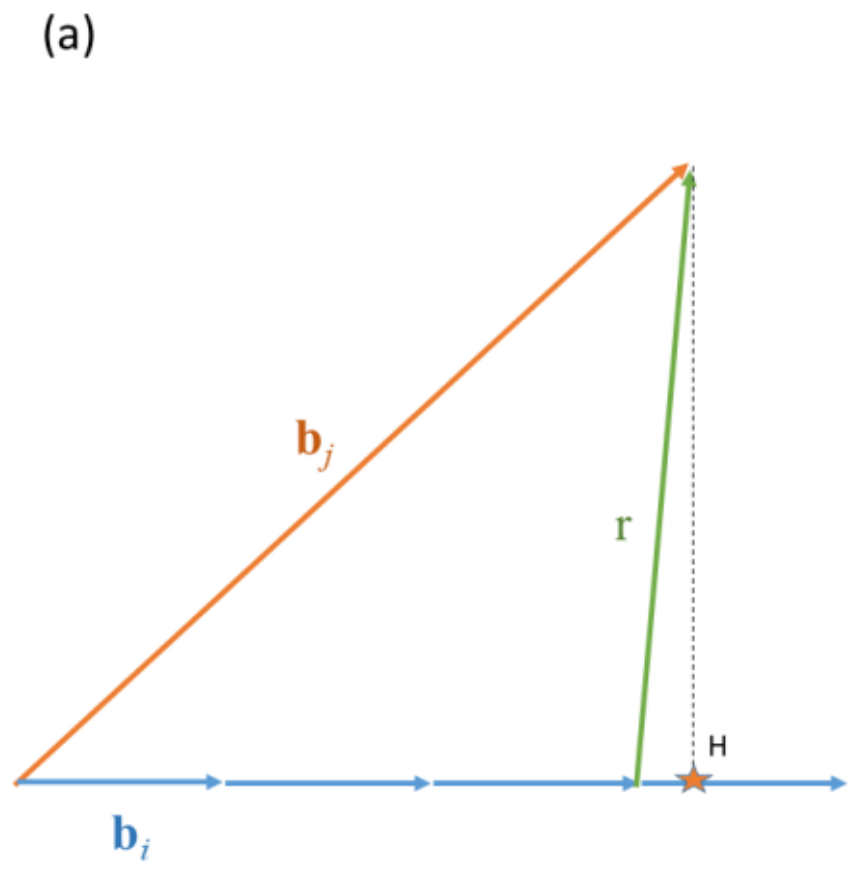

**Figure 2** Directional reduction. Case where (a) $\lfloor q \rceil = \left\lfloor \frac{\mathbf{b}_i^t \mathbf{b}_j}{\mathbf{b}_i^t \mathbf{b}_i} \right\rceil = 3$, and (b) $\lfloor q \rceil = 1$.The orthogonal projection point is noted H and marked by a little orange star.

The variant *Insert* gives good results in very short time. The deviation from cubicity and the sum of the squares of the norms of the list in Figure 1a that were initially $R = 453988268$ , $S = 61580172$ are reduced down to $R = 540$ , $S = 134$. These values are not far from those obtained with the LLL algorithm, i.e. $R = 531$ , $S = 99$. However, as it will be discussed in §4, the good efficiency of the variant *Insert* is made at the detriment of the hyperplanar reduction because of the "anisotropy" it generates. The variant *Append* gives better results than *Insert* for high dimensions, approximatively $N \geq 15$. The values for the list in Figure 1a are reduced "only" to $R = 1199$ , $S = 337$, but this will leave more action for the hyperplanar reduction, and better final reduction, as it will be shown in the next sections.

### 2.2. Simplification

Lagrange's division reduces the vectors by pairs without considering the basis as a whole. Now, if one accepts to possibly slightly but only temporarily degrades the value of S of the basis, the deviation from cubicity $R$ can be further decreased as follows. Let us consider again a list of integer vectors $\{\mathbf{b}_1, \ldots, \mathbf{b}_i, \ldots, \mathbf{b}_N\}$ sorted by norms from the lowest to the highest norms. For the pairs of vectors $\mathbf{b}_i$ and $\mathbf{b}_j$ in the list such that $\|\mathbf{b}_i\| \leq \|\mathbf{b}_j\|$, we calculate the vector $\mathbf{r} = \mathbf{b}_j - sign(\mathbf{b}_i^t \mathbf{b}_j)\, \mathbf{b}_i$ , where $sign(\mathbf{b}_i^t \mathbf{b}_j) = 1$ if $\mathbf{b}_i^t \mathbf{b}_j > 0$ , $-1$ if $\mathbf{b}_i^t \mathbf{b}_j < 0$ , and 0 if $\mathbf{b}_i^t \mathbf{b}_j = 0$. Then, we calculate whether or not replacing $\mathbf{b}_i$ or $\mathbf{b}_j$ by **r** allows to decrease the value of $R$. If the answer the positive, the change is made, and here again, two algorithm variants are proposed

> Variant "*Append*": If replacing $\mathbf{b}_i$ by **r** allows to decrease the value of $R$, the vector $\mathbf{b}_i$ is deleted and the vector **r** is appended at the end of the list. If not, the vector $\mathbf{b}_j$ is deleted and the vector **r** is appended at the end of the list.

> Variant "*Insert*": If replacing $\mathbf{b}_i$ by **r** allows to decrease the value of $R$, the vector $\mathbf{b}_i$ is replaced by **r** at the its position *i*, else, $\mathbf{b}_j$ is replaced by **r** at the its position *j*. The new list of vectors is then sorted again following the increasing norms.

The variant "*Insert*" is chosen by default, except for random matrices for which the variant "*Append*" should be preferred, as it will be discussed in §4. The process of simplification is repeated recursively until $R$ cannot be reduced anymore. Simplification permits to decrease a little more the values obtained in §2.1 for the list of Figure 1a, i.e. from $R = 1199$, $S = 337$ down to $R = 1084$, $S = 330$ with the variant *Insert*. At this step, no more improvement could be obtained, even by combining Lagrange's division and simplification. We will call "directional reduction" the result of Lagrange's division followed by simplification.

## 3. Hyperplanar reduction

### 3.1. The hyperplane normal

Let us consider again a list of integer vectors $\{\mathbf{b}_1, \ldots, \mathbf{b}_j, \ldots, \mathbf{b}_N\}$ initially sorted by norms, i.e. such that $\|\mathbf{b}_1\| \leq \cdots \leq \|\mathbf{b}_i\| \leq \|\mathbf{b}_{i+1}\| \ldots \leq \|\mathbf{b}_N\|$. We isolate the first vector $\mathbf{b}_1$ from the subspace of dimension $N-1$ (hyperplane) constituted of the vectors $\{\mathbf{b}_2, \ldots, \mathbf{b}_j, \ldots, \mathbf{b}_N\}$. The coordinates of the integer vector $\mathbf{p}_1$ that is normal to this hyperplane can be calculated as follows. We write the coordinates of vectors $\mathbf{b}_2, \ldots, \mathbf{b}_j, \ldots, \mathbf{b}_N$ in columns to form the $N$ x $N$-1 matrix

$$\mathbf{S}_1 = \begin{bmatrix} \mathrm{b}_{1,2} & \ldots & \mathrm{b}_{1,j} & \ldots & \mathrm{b}_{1,N} \\ \vdots & \ldots & \vdots & \ldots & \vdots \\ \mathrm{b}_{i,2} & \ldots & \mathrm{b}_{i,j} & \ldots & \mathrm{b}_{i,N} \\ \vdots & \ldots & \vdots & \ldots & \vdots \\ \mathrm{b}_{N,2} & \ldots & \mathrm{b}_{N,j} & \ldots & \mathrm{b}_{N,N} \end{bmatrix}$$

, where $\mathrm{b}_{i,j}$ means the $i^{\text{th}}$ coordinate of the vector $\mathbf{b}_j$.

If we insert as a first column any vector of the set $\{\mathbf{b}_2, \ldots, \mathbf{b}_j, \ldots, \mathbf{b}_N\}$, let us say the vector $\mathbf{b}_j$, then the new set of vectors becomes linearly dependent and the determinant of the $N$ x $N$ matrix is null.

$det\begin{bmatrix} \mathrm{b}_{1,j} & \mathrm{b}_{1,2} & \dots & \mathrm{b}_{1,j} & \dots & \mathrm{b}_{1,N} \\ \vdots & \vdots & \dots & \vdots & \dots & \vdots \\ \mathrm{b}_{i,j} & \mathrm{b}_{i,2} & \dots & \mathrm{b}_{i,j} & \dots & \mathrm{b}_{i,N} \\ \vdots & \vdots & \dots & \vdots & \dots & \vdots \\ \mathrm{b}_{N,j} & \mathrm{b}_{N,2} & \dots & \mathrm{b}_{N,j} & \dots & \mathrm{b}_{N,N} \end{bmatrix} = 0$ . Let us write this determinant by its cofactor expansion along the first column. The minors, i.e. the determinants of $\mathbf{M}_{1,k}$ , the $N-1 \times N-1$ submatrices of $\mathbf{S}_1$obtained by deleting the $k^{\text{th}}$ row, form a vector $\mathbf{p}_1 = \begin{bmatrix} +\,det(\mathbf{M}_{1,1}) \\ -\,det(\mathbf{M}_{1,2}) \\ \vdots \\ (-1)^{k+1}\, det(\mathbf{M}_{1,k}) \\ \vdots \\ (-1)^{N+1}\, det(\mathbf{M}_{1,N}) \end{bmatrix}$ that checks the property $\mathbf{p}_1^{\mathrm{t}}\mathbf{b}_j = 0, \forall j \in [2, \dots, N]$. In other words, $\mathbf{p}_1$ is the normal to the hyperplane $\{\mathbf{b}_2, \dots, \mathbf{b}_j, \dots, \mathbf{b}_N\}$ that we were looking for. Its norm equals the (hyper)surface formed by the vectors $\{\mathbf{b}_2, \dots, \mathbf{b}_j, \dots, \mathbf{b}_N\}$. The reader can check that in 3 dimension, $\mathbf{p}_1 = \mathbf{b}_2 \wedge \mathbf{b}_3$.

### 3.2. Reduction by orthogonal projection on the first hyperplane layer

Beside the metric tensor used to quantify the lattice rhombicity, we need another crystallographic concept, the “layers”. We will also employ the left inverse of matrices as coordinate transformation matrices between $N \times N$ space and its $N \times (N-1)$ subspaces, a tool we already introduced for $3 \times 3$ spaces in (Cayron, 2021a). Let us consider a unit cell of the lattice $\mathcal{L} = \{\mathbf{b}_1, \dots, \mathbf{b}_j, \dots, \mathbf{b}_N\}$ attached to the hyperplane $\{\mathbf{b}_2, \dots, \mathbf{b}_j, \dots, \mathbf{b}_N\}$. There are many equivalent unit cells, but we are looking for a quasi-reduced one. First we replace the $N-1$ sublattice $\{\mathbf{b}_2, \dots, \mathbf{b}_j, \dots, \mathbf{b}_N\}$ by its reduced form $\{\mathbf{b}'_2, \dots, \mathbf{b}'_j, \dots, \mathbf{b}'_N\}$ obtained by Lagrange’s division as described in §2.1. If the $N-1$ reduction is not possible, the sublattice $\{\mathbf{b}_2, \dots, \mathbf{b}_j, \dots, \mathbf{b}_N\}$ is not replaced. All the vectors $\mathbf{b}'_j$ belong to the hyperplane; we say that they are in the layer $q = 0$ of the plane $\mathbf{p}_1$. The vector $\mathbf{b}_1$ is the layer $q = 1$ of the hyperplane $\mathbf{p}_1$. The set $\{\mathbf{b}_1, \mathbf{b}'_2, \dots, \mathbf{b}'_j, \dots, \mathbf{b}'_N\}$ is a unit cell attached to the hyperplane. Vectors of the lattice $\mathcal{L}$ on the layer $q = 1$ shorten than $\mathbf{b}_1$ can be determined as follows. We note O the origin of the lattice and Z the point such that $\mathbf{OZ} = \mathbf{b}_1$, as illustrated in Figure 3. We call H the projection of O on the layer $q = 1$ of the hyperplane $\mathbf{p}_1$. It is such that $\mathbf{OH} \parallel \mathbf{p}_1$ and $\mathbf{p}_1^{\mathrm{t}}\,\mathbf{OH} = \mathbf{p}_1^{\mathrm{t}}\,\mathbf{b}_1$. Thus, $\mathbf{OH} = \left(\frac{\mathbf{p}_1^{\mathrm{t}}\,\mathbf{b}_1}{\mathbf{p}_1^{\mathrm{t}}\,\mathbf{p}_1}\right)\mathbf{p}_1$. Its coordinates are not integer but remain rational.

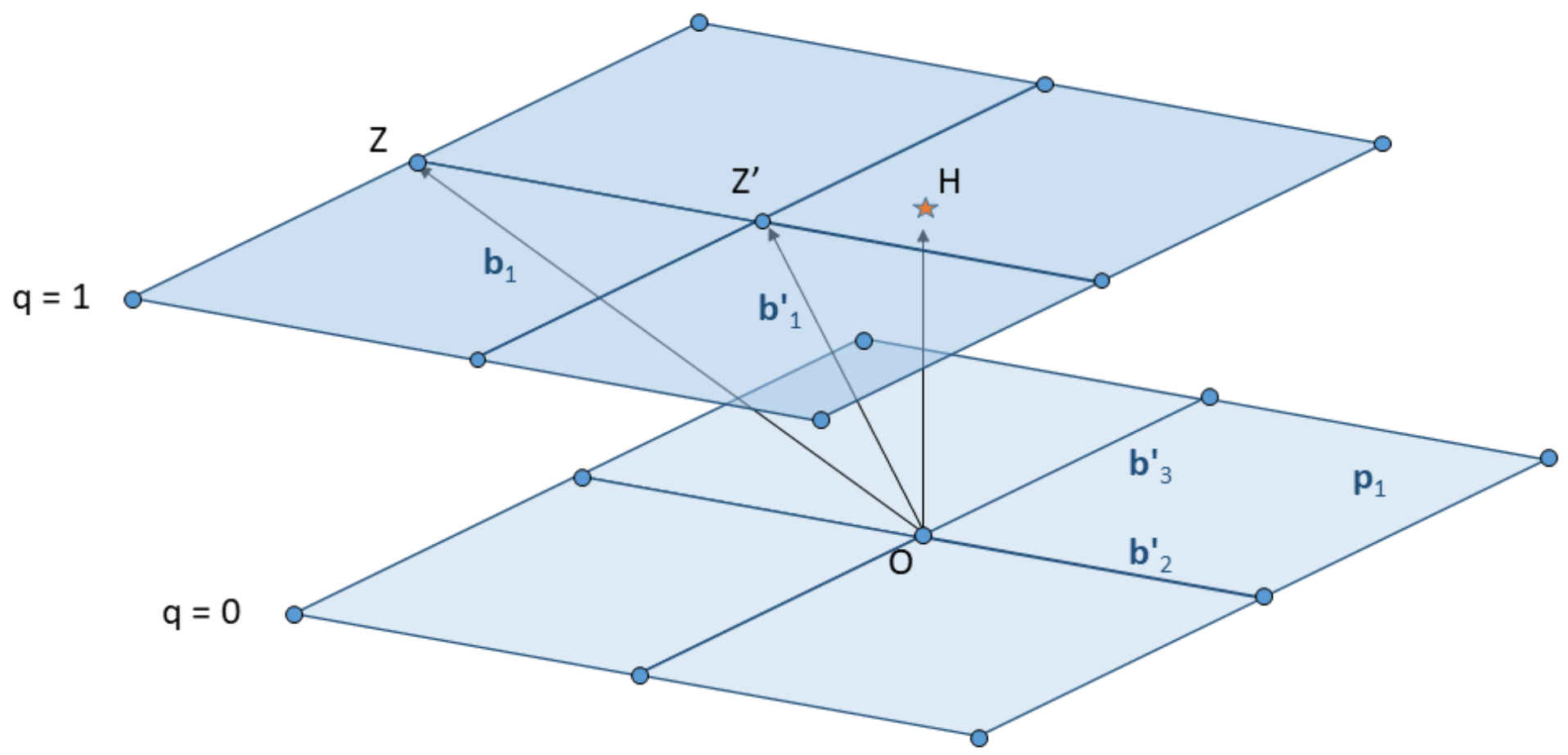

**Figure 3** Hyperplanar reduction. The lattice is "stratified" into different layers. Here only the layers $q = 0$ and $q = 1$ are represented. The vector of the layer $q = 1$ that should shorten/reduced is $\mathbf{OZ} = \mathbf{b}_1$. The reduction is made thanks to the calculation of the point H (marked by a little orange star) which is the orthogonal projection of the origin 0 onto the layer $q = 1$.

The vector $\mathbf{ZH} = -\mathbf{OZ} + \mathbf{ZH}$ is a vector of the hyperplane $\mathbf{p}_1$, which means that it can be written as a linear combination of the vectors $\{\mathbf{b}'_2, \dots, \mathbf{b}'_j, \dots, \mathbf{b}'_N\}$. In order to get its coordinates, we use again the $N$ x $N$-1 matrix formed by writing the reduced vectors in columns, i.e.

$$\mathbf{S}'_1 = \begin{bmatrix} \mathrm{b}'_{1,2} & \dots & \mathrm{b}'_{1,j} & \dots & \mathrm{b}'_{1,N} \\ \vdots & \dots & \vdots & \dots & \vdots \\ \mathrm{b}'_{i,2} & \dots & \mathrm{b}'_{i,j} & \dots & \mathrm{b}'_{i,N} \\ \vdots & \dots & \vdots & \dots & \vdots \\ \mathrm{b}'_{N,2} & \dots & \mathrm{b}'_{N,j} & \dots & \mathrm{b}'_{N,N} \end{bmatrix}.$$

The $N-1$ local coordinates of $\mathbf{ZH}$ in the basis $\{\mathbf{b}'_2, \dots, \mathbf{b}'_j, \dots, \mathbf{b}'_N\}$ are given by $\mathbf{ZH}_{loc} = (\mathbf{S}'_1)^{-1}_{Left} . \mathbf{ZH}$ where $(\mathbf{S}'_1)^{-1}_{Left}$ is the left inverse of the matrix $\mathbf{S}'_1$. We already used it in (Cayron, 2021); we recall that a left inverse of a non-square matrix $\mathbf{M}$ is $\mathbf{M}^{-1}_{\mathrm{Left}} = (\mathbf{M}^{\mathrm{t}}\, \mathbf{M})^{-1}\mathbf{M}^{\mathrm{t}}$. The vector $\mathbf{ZH}_{loc} = \{z_2, z_3, \dots, z_N\}$ is a rational $N-1$ dimensional vector in the $N-1$ subspace. A lattice point Z' close to H that belongs to the same layer is given by $\mathbf{Z'H}_{loc} = \{\lfloor z_2 \rceil, \lfloor z_3 \rceil, \dots, \lfloor z_N \rceil\}$. The vector $\mathbf{ZZ'}_{loc} = \mathbf{ZH}_{loc} - \mathbf{Z'H}_{loc}$ is calculated and re-expressed in the $N$-dimensional space by $\mathbf{ZZ'} = \mathbf{S}'_1 . \mathbf{ZZ'}_{loc}$. The vector $\mathbf{b}'_1 = \mathbf{OZ'} = \mathbf{OZ} + \mathbf{ZZ'}$ is a reduced form of the vector $\mathbf{b}_1$ . The rhombicity is re-calculated; if it has decreased, the solution $\{\mathbf{b}'_2, \dots, \mathbf{b}'_j, \dots, \mathbf{b}'_N, \mathbf{b}'_1\}$ replaces $\{\mathbf{b}_1, \dots, \mathbf{b}_j, \dots, \mathbf{b}_N\}$ and the process is repeated with the new list by recursion. If it has not decreased, the next vector in the list, $\mathbf{b}_2$, is used to calculate $\mathbf{p}_2$ and $\mathbf{b}'_2$, etc. The process stops when all the vectors $\mathbf{b}_i$ of the list $\{\mathbf{b}_1, \dots, \mathbf{b}_j, \dots, \mathbf{b}_N\}$ are screened but none of the vector $\mathbf{b}'_i$ allows to reduce anymore the lattice rhombicity. In materials science, we would say that $\mathbf{b}'_i$ is obtained from $\mathbf{b}_i$ by "simple shearing" on the hyperplane $\mathbf{p}_i$ . In the paper, we will call this operation "hyperplanar reduction". The reader can notice that both Lagrange's division and hyperplanar reduction imply orthogonal projections followed by numerical rounding in which rational numbers are replaced by their closest integers. After hyperplanar reduction, the lattice that was previously directionally reduced at the end of §2.2 becomes even more reduced since its values decreased to $R = 451$ and $S = 113$. These values are not yet as good as those obtained by LLL, but we have not yet used with the interplay between directional and hyperplanar reductions.

## 4. Cubification methods by cycling direction and hyperplanar reductions

### 4.1. Methods and options

Directional and hyperplanar reductions can now be repeated in cycles until the lattice rhombicity cannot be decreased anymore. We call this cycling "cubification". There is not a unique way to perform a cubification as it can be started by the hyperplanar reduction or by directional reduction. It also depends on the algorithm variants chosen for Lagrange's division (§2.1) and the simplification (§2.2) of the direction reduction. By trial-and-error, we could identify two cubification methods (Table 1).

**Table 1** Two cubification methods. Their options are given in Table 2.

| Method 1: | Method 2: |
|---|---|
| Cubification (list, *opt.*): | Cubification (list, *opt.*): |
| newlist = Sort_by_norm (list) | newlist = Sort_by_norm (list) |
| newlist = Directional reduction (newlist, *opt.*) | newlist = Hyperplanar reduction (newlist) |
| newlist = Hyperplanar reduction (newlist) | newlist = Directional reduction (newlist, *opt.*) |
| **If** $R$ (newlist) < $R$ (list): | newlist = Hyperplanar reduction (newlist) |
| **Return** Cubification (newlist, *opt.*) | **If** $R$ (newlist) < $R$ (list): |
| **Else** **Return** list | **Return** Cubification (newlist, *opt.*) |
| | **Else** **Return** list |

The algorithm variant that should be chosen depends on the type of matrix formed by the set of vectors that should be reduced (Table 2). We call here "columnar matrix" a list of vectors whose matrix (the vectors are written in rows) contains many zeros, and at least one column with many non-null and generally high integer values. A typical example is the matrix given in Figure 1a. We noticed that for large matrices, i.e. in dimensions approximatively $N \geq 15$, Lagrange's division in its *Append* variant gives better results. We call "heterogeneous matrix", a matrix that contains many zeros, and at least one row and one column with many non-null and generally high integer values. We noticed that for some cases of large heterogeneous matrices, with approximatively $N \geq 15$, the first directional reduction goes beyond the recursion limit of our computer; when this happens, a first hyperplanar reduction without $N-1$ directional reduction solves the problem. We call "random matrix" a matrix whose values are randomly computed with integers between 0 and 100. Larger limits, for example 1000, in large random matrices $N \geq 15$ leads to too high integer values in intermediate calculations and error messages both in *OLLL* and *Cubification*. A columnar random matrix is an identity matrix in which the last column is replaced by random integers in the range 0-100. Columnar random matrices are classified here as random matrices and are treated with method 2.

**Table 2** Method with options to be used depending of the type of the list of vectors (square matrix). We consider "large" a matrix of dimension $N \geq 15$. For some large heterogeneous matrices a first step with hyperplanar reduction without $N-1$ directional reduction may be required before starting Method 1 as indicated in parenthesis .

| | | Variant for the Directional reduction | |
|---|---|---|---|
| **Type of list of vectors** | **Cubification method** | **Lagrange's division** | **Simplification** |
| Small columnar matrix | Method 1 | *Insert* | *Insert* |
| Large columnar matrix | | *Append* | *Insert* |
| Large heterogeneous matrix | ($N-1$ Hyperplanar reduction +) Method 1 | *Insert* | *Insert* |
| Random matrix | Method 2 | *Append* | *Append* |

### 4.2. Computer program and comparisons

We wrote a computer program called *Cubification* in Python 3.8 using the Numpy library to perform the matrix calculations (scalar products, matrix products, inverses etc.) and to generate the random numbers, vectors and matrices. All the results presented in the paper were obtained with a 6 years old laptop equipped with Intel(R) Core(TM)i7-4600 CPU 2.1 GHz, 64-bit Windows system, with a RAM of 8 GB. The recursion limit in our Python program has been fixed to 10'000. We compared the results obtained with our program with those obtained by the LLL method computed in Python3 by Yonashiro (2020) and called *OLLL*. All the OLLL calculations were made with $\alpha = 3/4$. For specific matrices, as that of Figure 1, we also used the function *ReduceLattice* of *Mathematica*. On this example we checked that *OLLL* and *Mathematica* gives the same result; the only difference is that the calculations are nearly instantaneous with *Mathematica* whereas they are quite long with *OLLL*. As both use the same algorithm principles, this observation shows that one cannot compare easily the execution times to estimate the efficiency of an algorithm. Part of slowness of OLLL probably comes from the fact that Python is an interpreted language. In the rest of the paper, the execution times will be compared only between our program *Cubification* and *OLLL* because both are written in Python.

### 4.3. Results on non-random matrices

The best results of *Cubification* are obtained with columnar non-random matrices. For them our algorithm not only equals LLL but it seems to outperform it. For small columnar matrices, the cubification algorithm gives the same results as LLL, and for large $N \geq 10 - 15$ matrices, our experience shows that it gives in general reduced lists with $R$ and $S$ values that are lower than with LLL. For example, the lattice of Figure 1a could be reduced in only 3 cycles such that the final lattice is characterized by $R = 308$ , $S = 88$,; these values are clearly lower than those obtained by the LLL ($R = 531$ , $S = 99$). The output list of vectors is given in Figure 4. One reason of the better cubicity is that the ±1 are more homogenously distributed among the 20 columns. Now, if one compare the execution time, the result was obtained in only 2.5 s with *Cubification*, whereas 71 s were required for *OLLL*. The lattice that was already reduced by LLL given in Figure 1b could even be reduced more with *Cubification* in 1.2 s with final values $R = 360$ , $S = 92$. It can be noted that these values are however not as good as those obtained from the initial matrix in Figure 1a.

$$
\begin{bmatrix}
0 & 0 & 1 & 0 & 0 & 0 & 0 & 0 & 0 & 1 & 0 & 0 & 0 & 0 & -1 & 0 & 0 & 0 & 0 & 0 \\
0 & 0 & 0 & 0 & 0 & 0 & 0 & 0 & 0 & 0 & 1 & 0 & 0 & 1 & 0 & 0 & -1 & 0 & 0 & -1 \\
-1 & 0 & 0 & 0 & 0 & 0 & 0 & 0 & 0 & 0 & 0 & 0 & 1 & 0 & 1 & 1 & 0 & 0 & 0 & 0 \\
0 & 1 & 0 & 0 & 0 & 0 & 1 & 0 & -1 & 1 & 0 & 0 & 0 & 0 & 0 & 0 & 0 & 0 & 0 & 0 \\
0 & 0 & -1 & 0 & 0 & 0 & -1 & 0 & 0 & 0 & 0 & 0 & 0 & 0 & 0 & 0 & 1 & 0 & 0 & 1 \\
1 & 0 & 0 & 1 & 1 & 0 & 0 & 0 & 0 & 0 & 0 & 0 & 0 & 0 & 0 & 0 & 0 & 0 & 1 & 0 \\
0 & 0 & 0 & 1 & 0 & 0 & 0 & 0 & 1 & 0 & 0 & -1 & -1 & 0 & 0 & 0 & 0 & 0 & 0 & 0 \\
0 & 0 & 0 & 0 & 0 & 0 & 0 & 0 & 0 & 0 & 0 & 0 & 1 & -1 & -1 & 0 & 0 & 1 & 0 & 0 \\
0 & 0 & 0 & 0 & 0 & 0 & -1 & 0 & 0 & 0 & 1 & 0 & -1 & 0 & 0 & 0 & 1 & 0 & 0 & 0 \\
0 & 0 & 0 & 1 & 0 & 0 & 0 & 0 & 0 & -1 & 0 & 0 & 1 & 0 & -1 & 0 & 0 & 0 & 0 & 0 \\
0 & 1 & -1 & 0 & 0 & 0 & 0 & 1 & 0 & 0 & 0 & 0 & 0 & 0 & -1 & 0 & 0 & 0 & 0 & 0 \\
1 & 0 & 0 & 0 & 0 & 0 & 0 & 0 & 0 & 0 & 0 & 0 & 0 & 1 & 0 & 1 & 0 & 0 & 0 & 1 \\
0 & 0 & 0 & 1 & 0 & 1 & 0 & 0 & 0 & 0 & 0 & -1 & 0 & -1 & 1 & 0 & 0 & 0 & 0 & 0 \\
0 & 0 & 1 & 0 & 0 & 0 & -1 & 1 & 0 & 0 & 0 & 0 & 0 & 0 & 0 & 0 & 0 & 1 & 0 & -1 \\
0 & 0 & -1 & 0 & 1 & 0 & 0 & 0 & -1 & 0 & 0 & 0 & 0 & 0 & 0 & 0 & 0 & 0 & 1 & -1 \\
0 & 0 & 0 & 0 & 0 & 0 & 0 & 1 & 0 & 0 & 0 & 0 & 0 & 0 & -1 & 0 & 0 & -1 & -1 & -1 \\
0 & 0 & 1 & 0 & 0 & 0 & 0 & 0 & -1 & 0 & 0 & 0 & 1 & 0 & 0 & -1 & 1 & 0 & 0 & 0 \\
0 & 0 & 1 & 1 & 0 & 0 & 0 & 0 & 0 & 0 & 0 & 0 & 0 & 1 & 0 & 1 & 1 & 0 & 0 & 0 \\
-1 & 0 & 0 & 0 & 0 & 0 & 0 & 0 & 0 & 0 & 1 & 0 & 0 & 0 & -1 & 0 & -1 & 0 & 0 & 1 \\
-1 & 1 & 0 & 0 & 0 & 0 & 0 & 0 & 1 & 0 & -1 & 0 & 1 & 1 & 0 & 0 & 0 & 0 & 0 & 0
\end{bmatrix}
$$

**Figure 4** Cubification by the method 1 of the lattice of Figure 1a. The vectors are written in rows, as in Figure 1. The reduced lattice has for values $R = 308$ , $S = 88$.

For heterogeneous matrices, we have tested only five $20 \times 20$ matrices, and all of them show that LLL and cubification gives similar results. An example with a $20 \times 20$ matrix is given in Figure S1.

### 4.4. Results on random matrices

One way to automatically compare the algorithms is to generate random matrices. We have tested the performances of *Cubification* (method 2) and *OLLL* programs on columnar random matrices and full random matrices. We have tested the program on matrices of dimensions $10 \times 10$, $12 \times 12$ and $14 \times 14$. Fifty matrices have been generated for each type. The performances are measured by the reduction factors $\frac{R\,(\text{input})}{R(\text{output})}$ and $\frac{S\,(input)}{S\,(output)}$. The highest the reduction factors, the better the algorithm. The results are given in Table 3.

**Table 3** Reduction factors and mean time obtained on columnar and full random matrices of dimensions $10 \times 10$, $12 \times 12$ and $14 \times 14$ by testing 50 matrices. The mean deviation estimated by various tests is for *OLLL* around ± 20 % for 10x10 matrix and it decreases down to ± 5% for $14 \times 14$ matrix. It seems to be 30 % larger for *Cubification*.

| Reduction factor | $\frac{R\,(\text{input})}{R(\text{output})}$ | $\frac{S\,(input)}{S\,(output)}$ | Mean time per matrix |
|---|---|---|---|
| | Columnar random matrices $10 \times 10$ | | |
| *OLLL* | 2780 | 1000 | 2.4 s |
| *Cubification* | 3600 | 1060 | 0.11 s |
| | Columnar random matrices $12 \times 12$ | | |
| *OLLL* | 3120 | 1060 | 5.5 s |
| *Cubification* | 4100 | 1090 | 0.33 s |
| | Columnar random matrices $14 \times 14$ | | |
| *OLLL* | 3630 | 1160 | 10.5 s |
| *Cubification* | 4370 | 1070 | 0.54 s |
| | Full random matrices $10 \times 10$ | | |
| *OLLL* | 14.3 | 5.2 | 2.2 s |
| *Cubification* | 16.9 | 5.4 | 0.22 s |
| | Full random matrices $12 \times 12$ | | |
| *OLLL* | 14.1 | 5.0 | 5.6 s |
| *Cubification* | 15.2 | 4.6 | 0.71 s |
| | Full random matrices $14 \times 14$ | | |
| *OLLL* | 13.6 | 4.7 | 12.7 s |
| *Cubification* | 14.3 | 4.1 | 1.30 s |

The reduction of the rhombicity is systematically better with *Cubification* than with *OLLL* algorithm. The norms seem however less reduced by *Cubification* for large full random matrices. By comparing the execution times, *Cubification* appears to be approximatively ten times faster than OLLL. As the two computer programs use the same language (Python3) we can suppose that the cubification algorithm is at least as time effective as the LLL one. Clearly, more studies involving specialists of computer programming are required to better evaluate and compare the advantages and limits of both algorithms. One can even imagine a larger algorithm in which both methods would work together.

## 5. Conclusion and perspectives

We propose an alternative approach to the LLL algorithm for the problem of lattice reduction. It is based on the complementary actions of directional reduction and hyperplanar reduction. The former implies projections on the basis vectors (1 dimensional subspaces) and the latter on the basis hyperplanes ($N - 1$ dimensional subspaces). The algorithm does not require the calculations of Gram-Schmidt bases. The driving parameter of the reduction is the lattice rhombicity, a parameter that encompasses the information on the norms and on the deviations from orthogonality of the basis vectors. The results are quite equivalent to LLL for random and heterogeneous matrices, and systematically better for columnar matrices. We wrote a computer program in Python3 called *Cubification* and we compared its execution time with another Python3 computer program called *OLLL*. Our program is approximatively ten times faster, which makes us think that cubification may be more effective than LLL. More studies are however required to confirm or infirm this assertion.

We foresee margins of progression for cubification. The two cubification methods described in §4.1 were determined by trial-and-error; better strategies to alternate the directional and hyperplanar reduction processes seem possible, for example by cross-calling the two processes without necessarily screening all the vectors in the basis. We will also try to generalize to the whole program the $N \to N - 1$ decrease of dimensions already used in the hyperplanar reduction with help left inverse matrices. Reducing efficiently a lattice requires a strategy of optimisation in the space of possible paths. We think that our work on the Taylor series of optimized functions applied to the function "rhombicity" could be a way to find an effective path in reasonable time. It should to be recalled however that, as LLL, cubification gives only a *good* solution, but not necessarily the *best* one. More theoretical mathematics are required to evaluate how far the good solutions are from the optima. The new approach proposed in the present paper may probably help to get a new point of view on this problem and improve our understanding of the nature of its complexity.

Lattice reduction is used in various technological fields such as computer science, cryptography, image processing, etc. Cubification may be beneficial in these domains. We take the opportunity of this paper to draw the attention of crytographers on crystallography, and more especially on the groupoid composition table formed by the variants (simple-cosets) and their operators (double-cosets). These tables were introduced for phase transformations in materials (Cayron, 2006) but could have application in cryptography once generalized to groups larger than the 32 crystallographic point groups.

**Acknowledgements** Prof. Roland Logé is warmly acknowledged for the freedom given to our research that sometimes goes beyond metallurgy.

**<u>Note:</u>** In a first time, the Python program *Cubification* will be made available to the reviewers of the present manuscript. If the manuscript is accepted for publication, the program and its sources will be deposited on github or will be freely available on demand.

# Supporting information

### S1. Example of reduction of an heterogeneous matrix

(a) Input (the coordinates of the vectors are written in rows)

$$
\begin{bmatrix}
67 & 84 & -97 & -33 & 421 & 909 & -23 & 1 & 11 & -361 & 124 & 891 & 456 & 743 & 34 & 999 & 412 & 769 & 45 & -853 \\
-11 & -84 & 237 & -333 & 42 & 919 & -18 & 99 & 11 & -36 & 14 & 81 & 56 & -543 & 340 & 99 & -48 & 76 & 145 & -313 \\
0 & 0 & 1 & 0 & 0 & 0 & 0 & 0 & 0 & 0 & 0 & 0 & 0 & 0 & 0 & 0 & 0 & 0 & 0 & 1586 \\
0 & 0 & 0 & 1 & 0 & 0 & 0 & 0 & 0 & 0 & 0 & 0 & 0 & 0 & 0 & 0 & 0 & 0 & 0 & 1030 \\
0 & 0 & 0 & 0 & 1 & 0 & 0 & 0 & 0 & 0 & 0 & 0 & 0 & 0 & 0 & 0 & 0 & 0 & 0 & 1921 \\
0 & 0 & 0 & 0 & 0 & 1 & 0 & 0 & 0 & 0 & 0 & 0 & 0 & 0 & 0 & 0 & 0 & 0 & 0 & 569 \\
0 & 0 & 0 & 0 & 0 & 0 & 1 & 0 & 0 & 0 & 0 & 0 & 0 & 0 & 0 & 0 & 0 & 0 & 0 & -721 \\
0 & 0 & 0 & 0 & 0 & 0 & 0 & 1 & 0 & 0 & 0 & 0 & 0 & 0 & 0 & 0 & 0 & 0 & 0 & 1183 \\
0 & 0 & 0 & 0 & 0 & 0 & 0 & 0 & 1 & 0 & 0 & 0 & 0 & 0 & 0 & 0 & 0 & 0 & 0 & 1570 \\
0 & 0 & 0 & 0 & 0 & 0 & 0 & 0 & 0 & 1 & 0 & 0 & 0 & 0 & 0 & 0 & 0 & 0 & 0 & -6665 \\
0 & 0 & 0 & 0 & 0 & 0 & 0 & 0 & 0 & 0 & 1 & 0 & 0 & 0 & 0 & 0 & 0 & 0 & 0 & 123 \\
0 & 0 & 0 & 0 & 0 & 0 & 0 & 0 & 0 & 0 & 0 & 1 & 0 & 0 & 0 & 0 & 0 & 0 & 0 & 890 \\
0 & 0 & 0 & 0 & 0 & 0 & 0 & 0 & 0 & 0 & 0 & 0 & 1 & 0 & 0 & 0 & 0 & 0 & 0 & 6 \\
0 & 0 & 0 & 0 & 0 & 0 & 0 & 0 & 0 & 0 & 0 & 0 & 0 & 1 & 0 & 0 & 0 & 0 & 0 & 742 \\
0 & 0 & 0 & 0 & 0 & 0 & 0 & 0 & 0 & 0 & 0 & 0 & 0 & 0 & 1 & 0 & 0 & 0 & 0 & 33 \\
0 & 0 & 0 & 0 & 0 & 0 & 0 & 0 & 0 & 0 & 0 & 0 & 0 & 0 & 0 & 1 & 0 & 0 & 0 & 888 \\
0 & 0 & 0 & 0 & 0 & 0 & 0 & 0 & 0 & 0 & 0 & 0 & 0 & 0 & 0 & 0 & 1 & 0 & 0 & 14 \\
0 & 0 & 0 & 0 & 0 & 0 & 0 & 0 & 0 & 0 & 0 & 0 & 0 & 0 & 0 & 0 & 0 & 1 & 0 & 769 \\
0 & 0 & 0 & 0 & 0 & 0 & 0 & 0 & 0 & 0 & 0 & 0 & 0 & 0 & 0 & 0 & 0 & 0 & 1 & 1234 \\
0 & 0 & 0 & 0 & 0 & 0 & 0 & 0 & 0 & 0 & 0 & 0 & 0 & 0 & 0 & 0 & 0 & 0 & 0 & -853
\end{bmatrix}
$$

(b) Output LLL

$$
\begin{bmatrix}
0 & 0 & 0 & -1 & 0 & 1 & 1 & 1 & 0 & 0 & 0 & 0 & 0 & 0 & 0 & 0 & 0 & 0 & 0 & 1 \\
0 & 0 & 0 & 1 & -1 & 0 & 0 & 0 & 0 & 0 & 0 & 1 & 0 & 0 & 0 & 0 & 0 & 0 & 0 & -1 \\
0 & 0 & -1 & 0 & 0 & 0 & -1 & 0 & 0 & 0 & 1 & 0 & 0 & 1 & 0 & 0 & 0 & 0 & 0 & 0 \\
0 & 0 & 0 & 0 & 0 & -1 & 0 & 0 & 1 & 0 & 0 & -1 & 0 & 1 & 0 & 0 & 0 & 0 & 0 & 0 \\
0 & 0 & 0 & 0 & 0 & 0 & -1 & 0 & -1 & 0 & 0 & -1 & 0 & 0 & 1 & 0 & 0 & 0 & 0 & 0 \\
0 & 0 & 1 & 0 & -1 & 1 & 0 & 0 & 0 & 0 & -1 & 0 & 0 & 1 & 0 & 0 & 0 & 0 & 0 & 0 \\
0 & 0 & -1 & 1 & 0 & 0 & 0 & -1 & 0 & 0 & 0 & 0 & 0 & 0 & 1 & 0 & 0 & 0 & 0 & 0 \\
0 & 0 & 0 & -1 & 0 & 0 & 1 & 0 & 1 & 0 & 0 & 1 & 0 & -1 & 1 & 0 & 0 & 0 & 0 & 0 \\
0 & 0 & 0 & -1 & 0 & 0 & 1 & 0 & 0 & 0 & 1 & 0 & 0 & 1 & 1 & 0 & 0 & 0 & 0 & 0 \\
0 & 0 & 1 & 0 & -1 & 0 & 0 & 1 & 0 & 0 & 0 & 0 & 1 & 0 & 0 & 0 & 0 & 0 & 0 & 1 \\
0 & 0 & 0 & -1 & 0 & 0 & 1 & 0 & 0 & 0 & 0 & 1 & -1 & 0 & 0 & 0 & 1 & 0 & 0 & 0 \\
0 & 0 & 0 & 0 & 0 & 0 & 1 & 0 & 0 & 0 & -1 & -1 & -1 & 0 & 0 & 1 & 0 & 0 & 0 & 1 \\
0 & 0 & 0 & 0 & 0 & 1 & 0 & 1 & 1 & 0 & 1 & 0 & 0 & 0 & -1 & 0 & 0 & 0 & 0 & 0 \\
0 & 0 & 0 & -1 & 0 & 0 & 1 & 0 & 0 & 0 & 0 & -1 & 0 & 0 & 1 & 1 & 1 & 0 & 0 & 0 \\
0 & 0 & 0 & 0 & 0 & -1 & 0 & -1 & 0 & 0 & 0 & 0 & 0 & -1 & 1 & 0 & -1 & 1 & 0 & 0 \\
0 & 0 & 1 & 0 & -1 & 0 & 0 & 0 & 0 & 0 & -1 & 0 & 0 & 0 & 0 & 0 & -1 & 0 & -1 & 0 \\
0 & 0 & -1 & 0 & -1 & 0 & 1 & 0 & 0 & 0 & 0 & -1 & 0 & 0 & 0 & 0 & 0 & 0 & 0 & 0 \\
0 & 0 & 0 & 0 & 0 & 0 & -1 & 0 & 0 & 1 & 0 & 0 & 1 & 0 & -1 & 0 & 0 & 0 & 0 & 0 \\
56 & 0 & 0 & 0 & 0 & 0 & 0 & -1 & 0 & 0 & -1 & -1 & 0 & 0 & 0 & 0 & 0 & 0 & 0 & 0 \\
-11 & -84 & 0 & 0 & 0 & 1 & 0 & 1 & 0 & 0 & 0 & 0 & 0 & 1 & -1 & 0 & 0 & 0 & 0 & 0
\end{bmatrix}
$$

(c) Output cubification (method 1)

$$\begin{bmatrix}
0 & 0 & 0 & 0 & 0 & 0 & 0 & 0 & -1 & -1 & 0 & 1 & 0 & 0 & 0 & 0 & -1 & 0 & 0 & 0 \\
0 & 0 & 0 & 0 & 0 & -1 & 0 & 0 & 1 & 0 & 0 & -1 & 0 & 1 & 0 & 0 & 0 & 0 & 0 & 0 \\
0 & 0 & 0 & 0 & 0 & 0 & 0 & 0 & 0 & 0 & 0 & 1 & 0 & 0 & 1 & 0 & 1 & 1 & 0 & 0 \\
0 & 0 & -1 & 0 & 0 & 0 & 0 & 0 & 0 & 0 & 0 & -1 & 0 & 0 & 0 & 0 & 0 & 1 & 0 & -1 \\
0 & 0 & -1 & 0 & 0 & 0 & 0 & 0 & 0 & -1 & 0 & 0 & 1 & 0 & 1 & 0 & 0 & 0 & 0 & 0 \\
0 & 0 & 0 & -1 & 0 & 0 & 0 & 0 & -1 & 0 & 0 & 0 & 1 & 0 & 0 & 1 & 0 & 0 & 0 & 0 \\
0 & 0 & 0 & 0 & 1 & 0 & -1 & 0 & 0 & 0 & 0 & 0 & 0 & 0 & 0 & 0 & 0 & 1 & 0 & -1 \\
0 & 0 & 0 & 1 & 0 & -1 & -1 & -1 & 0 & 0 & 0 & 0 & 0 & 0 & 0 & 0 & 0 & 0 & 0 & -1 \\
0 & 0 & 0 & 0 & -1 & 1 & 0 & 0 & 0 & -1 & 0 & 0 & 0 & 1 & 0 & 0 & 0 & 1 & 0 & 0 \\
0 & 0 & 0 & 0 & 0 & -1 & 0 & 0 & -1 & 0 & -1 & 0 & 0 & 0 & 0 & 0 & 0 & -1 & -1 & 0 \\
0 & 0 & 0 & 0 & 1 & 0 & 0 & 0 & 1 & 0 & 0 & 0 & 1 & 0 & 0 & 0 & 0 & 1 & 0 & 1 \\
0 & 0 & 0 & 0 & 1 & 0 & 0 & 0 & 0 & 0 & -1 & 0 & 1 & 0 & 0 & 0 & -1 & 1 & 0 & 0 \\
0 & 0 & 0 & -1 & 0 & 0 & -1 & 1 & 0 & 0 & 0 & 0 & 0 & 0 & 0 & -1 & 1 & 0 & 0 & 0 \\
0 & 0 & 0 & 0 & 0 & 0 & 0 & 0 & 0 & 0 & 1 & 1 & 0 & 1 & 0 & 1 & 0 & 1 & 0 & 0 \\
0 & 0 & 0 & -1 & 0 & -1 & -1 & 0 & -1 & 0 & 0 & 0 & 0 & 1 & 0 & 0 & 0 & 0 & 0 & 0 \\
0 & 0 & 0 & 0 & 1 & 0 & 1 & 0 & 0 & 0 & 0 & 0 & 0 & 0 & 1 & 0 & 0 & 0 & -1 & -1 \\
0 & 0 & 1 & 1 & 0 & 0 & 0 & 0 & -1 & 0 & -1 & 0 & 0 & 0 & 0 & 0 & 1 & 1 & 0 & 0 \\
0 & 0 & 0 & -1 & 1 & 1 & 1 & 0 & -1 & 0 & 0 & -1 & 0 & 0 & 0 & 0 & 1 & 0 & 0 & -1 \\
-56 & 0 & 0 & 0 & -1 & 0 & 0 & 0 & 0 & 0 & -1 & 0 & 0 & -1 & 0 & 0 & 0 & 0 & 0 & -1 \\
-11 & -84 & 0 & 0 & 0 & 0 & -1 & 0 & 0 & 0 & 0 & 0 & 0 & 0 & 1 & 0 & 0 & 0 & 0 & -1
\end{bmatrix}$$

**Figure S1** Example of reduction of heterogeneous matrix with LLL and cubification. (a) Input matrix $R = 489734657$, $S = 68191151$ , (b) matrix reduced by LLL, $R = 12007$, $S = 10407$, (c) matrix reduced by cubification $R = 11905$, $S = 10407$.